\documentclass[aps,prb,twocolumn,showpacs,showkeys]{revtex4-1}

\usepackage{graphicx}
\usepackage{mathptmx}      
\usepackage{amsmath}
\usepackage{amsfonts}

\newcommand{\qop}[1]{\hat{#1}}
\newcommand{\hefour}{\textsuperscript{4}He}

\usepackage{bm}

\usepackage{graphicx}
\usepackage{color}

\begin{document}


\title{A coordinated wavefunction for the ground state of liquid \hefour}


\author{Y. Lutsyshyn}
\homepage[QTMPS group: ]{http://www.physik.uni-rostock.de/qtmps/\\}
\email[email:]{yaroslav.lutsyshyn@uni-rostock.de}
\affiliation{Institut f\"{u}r Physik, Universit\"{a}t Rostock, 18051 Rostock, Germany}


\date{\today}

\begin{abstract}
We present a variational ansatz for the ground state of a strongly correlated Bose system. This ansatz goes beyond the Jastrow-Feenberg functional form and explicitly enforces coordination shells in the structure of the wavefunction. We apply this ansatz to liquid helium-4 with a simple three-variable parametrization of the pair functions. The optimized wavefunction is found to give an excellent description of the mid-range correlations in the fluid. We also demonstrate the possibility to use this ansatz to study inhomogeneous systems. The phase separation and free surface emerge naturally in this wavefunction, even though it is constructed of short-range two-body functions and does not contain one-body terms. Because no explicit description of the surface is necessary, this provides a powerful description tool for cluster states.
\end{abstract}

%
%

\maketitle




\section{Introduction}

The interest in the microscopic nature of the ground state of liquid \hefour\ has drawn attention
for over half a century and has shaped the development of many aspects of the quantum many-body theory \cite{RPMBT-Volume5}.
The question continues to be on interest, especially as new correlated bosonic 
systems are becoming the subject of an experiment, including the cold atomic gases %
\cite{Trippenbach2012-ColdAndTrappedMetastableNobleGases,%
Ketterle2002-NobelLecture,%
Greywall1993-HeatCapacityAndTheCommensurateIncommensurateTransitionOfHe4AdsorbedOnGraphite,%
Kyotani2011-PhaseDiagramOf4HeFilmIn3DNanoporesOfZTC}.
An explicit and numerically efficient expression for the many-body wavefunction 
also has a practical use in computer calculations.
%
%
A good approximation to the ground state reduces the numerical costs and improves the statistical accuracy of the 
true ground state results
obtained with the diffusion Monte Carlo 
   \cite{Anderson1980,Ceperley1980} 
as well as the path-integral ground state Monte Carlo 
   \cite{Schmidt2000-APathIntegralGroundStateMethod,%
   Reatto2003-RecentProgressInSimulationOfTheGroundStateOfManyBosonSystems,%
   Boninsegni2005-PathIntegralGroundStateWithAFourthOrderPropagatorApplicationToCondensedHelium,%
   Boronat2010-HighOrderTimeExpansionPathIntegralGroundState%
}
methods. 

The variational ansatz for liquid \hefour\ has followed the path of improving  the Jastrow-Feenberg form of the wavefunction
\cite{Feenberg1974-GroundStateOfAnInteractingBosonSystem,%
Campbell1977-EnergyAndStructureOfTheGroundStateOfLiquidHeFour},
\begin{equation}
\psi(\bm{r}_1,\dots,\bm{r}_N) = \prod_{i<j}e^{\frac12 u_2(\bm{r}_i-\bm{r}_j)} 
\prod_{i<j<k}e^{\frac12 u_3(\bm{r}_i-\bm{r}_j,\bm{r}_j-\bm{r}_k,\bm{r}_k-\bm{r}_i)} \dots ,
\label{eq:FeenbergForm}
\end{equation}
where $N$ is the number of atoms, and the $k$-body correlation 
factors $u_k$ must have proper symmetry under the exchange of particles.
Because each successive term  in Eq.~(\ref{eq:FeenbergForm}) increases the
the numerical complexity by an additional factor of $N$, one is in practice limited to two- and three-body terms.
Limiting Eq.~(\ref{eq:FeenbergForm}) to two-body factors results in the Jastrow function \cite{Bijl,Jastrow}.
In an early work, McMillan \cite{McMillan} and Schiff and Verlet \cite{SchiffVerlet-1967}
used a Jastrow function with the two-body function 
$
u_2 = -{\left({b}/{r}\right)}^5
$.
Parameter $b$ was determined variationally.
The McMillan function captures the most significant
features of the system  caused by the core of the interparticle potential 
and it continues to be used successfully as a guiding function for projector Monte Carlo%
   \cite{%
      Boronat2010-HighOrderTimeExpansionPathIntegralGroundState,%
      Boronat2005-QuantumMonteCarloSimulationOfOverpressurizedLiquidHe4,%
      Cazorla2012-TheElasticConstantsOfSolid4HeUnderPressureADiffusionMonteCarloStudy%
   }.
Successive improvements in the ground state of helium  refined the two- and, later, three- body factors in the form~(\ref{eq:FeenbergForm}).
Published progress on this topic is too numerous to cover in any detail here. 
Relevant to this work, we note the addition
of the mid-range correlation
\cite{Reatto1974-HowGoodCanJastrowWavefunctionsBeForLiquidHeliumFour,%
Reatto1979-SpatialCorrelationsAndElementaryExcitationsInManyBodySystems}
which among other things allowed to replicate the first correlation peak of the pair distribution function $g(r)$;
the addition of long-range terms in the two-body function $u_2$ that allows to account for the long-wavelength 
zero-point phonons 
\cite{ReattoChester1967-PhononsAndThePropertiesOfABoseSystem,%
Reatto1970-GroundStateOfLiquidHeFour};
the computation of $u_2$ based on the maximum overlap with the true ground-state 
\cite{Reatto1984-SolutionOfAnInverseProblemMaximumOverlapJastrowFunctionOfTheLennardJonesBoseFluid,%
Reatto1987-MaximumOverlapJastrowWaveFunctionForLiquidHeFour}; 
and finally, a detailed optimization 
of the pair factors expanded in terms of the pair scattering eigenstates 
\cite{Vitiello1992-OptimizationOfHeWaveFunctionsForTheLiquidAndSolidPhases,%
Vitiello1999-VariationalMethodsForHeUsingAModernHeHePotential} which
 along with the inclusion of the three-body factors allowed
to account for nearly all the correlation energy.
The success of the above works came at the expense of the increased complexity and 
and the number of variational parameters that are needed to accurately 
describe the functions $u_k$. The general functional form of Eq.~(\ref{eq:FeenbergForm}), 
though, remained unchanged%
\footnote{A prominent exception, the Feynman-Cohen backflow wavefunction
\cite{Feynman1956-EnergySpectrumOfTheExcitationsInLiquidHelium}, is in fact not designed
for a ground state of bosonic many-body system, but is instead commonly used for the excited states of \hefour\
\cite{Feynman1956-EnergySpectrumOfTheExcitationsInLiquidHelium,%
Reatto2014-QuantumMonteCarloStudyOfAVortexInSuperfluidHe4AndSearchForAVortexStateInTheSolid,%
Boronat2005-QuantumMonteCarloSimulationOfOverpressurizedLiquidHe4}
and for the 
fermionic 
systems 
\cite{Carlson1989-FermionMonteCarloAlgorithmsAndLiquidHe3,%
Ceperley1993-EffectsOfThreeBodyAndBackflowCorrelationsInTheTwoDimensionalElectronGas,%
Boronat2012-FerromagneticTransitionOfATwoComponentFermiGasOfHardSpheres,%
Needs2013-QuantumMonteCarloCalculationOfTheFermiLiquidParametersOfTheTwoDimensionalHomogeneousElectronGas,%
Galli2015-ImplementationOfTheLinearMethodForTheOptimizationOfJastrowFeenbergAndBackflowCorrelations}.
}.

The development of the shadow wavefunction (SWF) methods 
   \cite{Vitiello1988-VariationalCalculationsForSolidAndLiquidHe4WithAShadowWaveFunction,%
   Reatto1988-ShadowWaveFunctionForManyBosonSystems} 
has to a large degree overtaken the development of the wavefunction for liquid helium.
The SWF allows to account for the correlations missed by the Jastrow function,
and results in an excellent description in terms of both energy and structure
   \cite{Reatto1992-AStudyOfTheLiquidPhaseOf4HeUsingAnImprovedShadowWaveFunction,%
   Reatto1994-TrialShadowWaveFunctionForTheGroundStateOfHe4}
of \hefour.
Relevant to this work, we notice that SWF can support self-bound states of liquid \hefour\
   \cite{Reatto1994-ASelfBoundWavefunctionForClustersOfHe4}.
Shadow wavefunction accounts for correlations via 
integrals on auxiliary (shadow) variables. 
The inclusion of the shadow variables may be seen as going beyond 
the Jastrow-Feenberg form of Eq.~(\ref{eq:FeenbergForm}).
However, the integrals on the shadow variables must be taken numerically by a Monte Carlo scheme, 
and in this sense shadow wavefunction
is not explicit, partially limiting its adoption in quantum Monte Carlo.

We will present a variational ansatz for the ground state of liquid \hefour\ which 
is build upon the Jastrow wavefunction but goes beyond  the general functional form of Eq.~(\ref{eq:FeenbergForm}).
This ansatz allows to explicitly control the mid-range structure of the liquid 
and results in a stark improvement of the atomic pair distribution already with a three-parameter wavefunction.
The wavefunction is presented in Section \ref{sec:Ansatz} and the computational results are shown
in Section \ref{sec:Results}. Section \ref{sec:SelfBound} presents results for inhomogeneous systems, followed by a discussion.

\section{The coordinated wavefunction \label{sec:Ansatz}}

\subsection{Variational ansatz}

Our proposed wavefunction consists of a product of the Jastrow function (limiting Eq.~(\ref{eq:FeenbergForm}) to two-body terms) and of the additional term to which we refer as the ``coordination term''.
The wavefunction has the following form,
\begin{equation}
\psi_{JC}(\bm{r}_1,\dots, \bm{r}_N) = \prod_{i<j}e^{\frac12 u_2(|\bm{r}_i-\bm{r}_j|)} \prod_{i}\sum_{j\ne i} y_2(|\bm{r}_i-\bm{r}_j|).
\label{eq:CoordinatedWavefunction}
\end{equation}
The factors $y_2(r)$ must vanish at large distances.
At short distances, $y_2$ is expected to raise to a constant.

  
The effect of the coordination term in (\ref{eq:CoordinatedWavefunction}) can be seen by 
inspection. Suppose the function $y_2(r)$ vanishes for distances $r$ beyond the mean interparticle distance. 
In this case, $y_2(r_{ij})$ will have significant value only for the pairs of immediate neighbors $\langle i,j\rangle$. 
On the other hand, the number of neighbors for each atom is limited by the presence of the repulsive core and by the Jastrow part of the wavefunction.
Thus the overall number of non-vanishing terms $y_2(r_{ij})$ in the system is, roughly speaking, fixed.
Under such a restraint, the product of sums in the coordination part of Eq.~(\ref{eq:CoordinatedWavefunction})
is maximized when all sums are equal to each other. 
That is, \emph{the non-vanishing values of  $y_2(r_{ij})$ are  distributed equally between the products}.
The wavefunction $\psi_{JC}$,
while constructed only of pairwise functions, has a ``global'' property in that it explicitly demands 
that each atom in the system has an equal expected number of immediate neighbors.
As we will see, this allows to improve the mid-range properties of the system independently of the Jastrow factor.

\subsection{Inspiration and origin}

The inspiration for the coordinated wavefunction $\psi_{JC}$ comes 
from the symmetrized Bose-solid wavefunction proposed by Cazorla et al.~\cite{Boronat2009NJP}.
%
This symmetrical solid wavefunction does an excellent 
work describing quantum Bose solid, both variationally
   \cite{Lutsyshyn2014-Phase}
and as a guiding function for importance sampling in quantum Monte Carlo simulations of Bose solids
   \cite{Lutsyshyn2010-InstabilityOfVacancyClustersInSolid4He,%
   Lutsyshyn2011-OnTheStabilityOfSmallVacancyClustersInSolid4He,%
   Lutsyshyn2010-PropertiesOfVacancyFormationInHcp4HeCrystalsAtZeroTemperatureAndFixedPressure,%
   Boronat2013-ElasticConstantsOfIncommensurateSolidHe4FromDiffusionMonteCarloSimulations}. 
In fact, one will recognize that Eq.~(\ref{eq:CoordinatedWavefunction}) 
is the wavefunction of Cazorla et al., except 
that the site locations of a crystalline structure are here replaced  by the positions of atoms themselves.

The solid wavefunction of Ref.~\onlinecite{Boronat2009NJP}  forces atoms to be located in the vicinity  
of one of the externally specified lattice sites, while at the same time imposing the global restraint
by favoring single site occupancy. 
In the liquid, the translational symmetry is not broken and there are no preferred positions;
instead, the atoms in (\ref{eq:CoordinatedWavefunction}) are ``localized'' around their neighbors.
As the overlap of atomic cores is prohibited by the Jastrow term, this creates the coordination shells.
  
An important distinction between $\psi_{JC}$ of Eq.~(\ref{eq:CoordinatedWavefunction})
and the symmetrized Nosanow-Jastrow wavefunction of Ref.~\onlinecite{Boronat2009NJP}
is in the nature of the sum-factors.
As discussed in Ref.~\onlinecite{Lutsyshyn2014-Phase}, factors that bind atoms to the lattice 
sites in the solid wavefunction of Ref.~\onlinecite{Boronat2009NJP} can be seen as a generalized symmetrical form of the one-body factor;
the coordination part of Eq.~(\ref{eq:CoordinatedWavefunction}), however, is by the same criterion a full $N$-body term.

\subsection{Separability}

If the particles of the system are divided into subgroups separated by a large distance, the wavefunction ought to reduce to the product of the wavefunctions for the individual subgroups \cite{QMBT-V7}. 
While such cluster property is obviously satisfied by the Jastrow function, it is less transparent for the coordination term. Suppose all particles are divided into two groups, or clusters, $A$ and $B$. Let the corresponding number of particles be $N_A$ and $N_B$, $N_A+N_B=N$. The distance between these clusters is sufficiently large such that the function $y_2$ vanishes for any pair of particles from across the two groups,
\begin{equation*}
\forall i\in A, j\in B: y_2(|\bm{r}_i-\bm{r}_j|)=0.
\end{equation*}
In this case, the coordination sum for any particle in a subgroup reduces 
to the sum on that subgroup only, and the  
coordination term separates,
\begin{multline*}
\psi_C(\bm{r}_1,\dots,\bm{r}_N) = \prod_i \sum_{j\ne i} y_2(|\bm{r}_i-\bm{r}_j|)  \\
   = \left(\prod_{i\in A}  \sum_{j\ne i} y_2(|\bm{r}_i-\bm{r}_j|) \right)
      \times  
      \left(\prod_{i\in B} \sum_{j\ne i} y_2(|\bm{r}_i-\bm{r}_j|)  \right) \\
   = \left(\prod_{i\in A}  \sum_{\substack{j\ne i \\  j\in A}} y_2(|\bm{r}_i-\bm{r}_j|) \right)
      \times  
      \left(\prod_{i\in B} \sum_{\substack{j\ne i \\  j\in B}} y_2(|\bm{r}_i-\bm{r}_j|)  \right) \\
    =
\psi_C(\underbrace{\bm{r}_{i'_1}, \dots,\bm{r}_{i'_{N_A}}}_{i'\in A}  )  \times 
\psi_C(\underbrace{\bm{r}_{i''_1}, \dots,\bm{r}_{i''_{N_B}}}_{i''\in B}     )  \\
 =   \psi_C(A) \psi_C(B).
\end{multline*}
Thus a wavefunction for the two clusters reduces to the product of the wavefunctions for the individual clusters.

\subsection{Computational complexity}

The evaluation of the  coordinated wavefunction of Eq.~(\ref{eq:CoordinatedWavefunction}) requires
the computation of ${O}(N^2)$ interparticle distances, and the overall computational cost also scales as the second 
order in the number of particles.
The scaling holds for the application of the Hamiltonian and other relevant operators.
To see this, we write Eq.~(\ref{eq:CoordinatedWavefunction}) as
\[
\psi_{JC}(\bm{r}_1,\dots,\bm{r}_N) = \prod_{i<j}e^{\frac12 u_2(|\bm{r}_i-\bm{r}_j|)} \prod_{i}  S_i(\bm{r}_1,\dots,\bm{r}_N),
\]
with
\[
S_i(\bm{r}_1,\dots,\bm{r}_N) = \sum_{j\ne i} y_2(|\bm{r}_i-\bm{r}_j|).
\]
In order to compute all $N$ sums $S_i$, one needs to compute $N (N-1)/2$ values of $y_2(|\bm{r}_i-\bm{r}_j|)$,
so long as the sums are stored in memory. 
This is not a taxing requirement, given one must in any case store $3N$ atomic coordinates.
Once the sums are computed, the computation of the product $\prod_{i} S_i$ only requires $N$ operations.

Similar considerations apply to the computation of the Hamiltonian and other relevant expressions.
For quantum Monte Carlo, one generally needs to compute the contribution to the local kinetic energy 
$\sum_i \frac{\nabla_i^2 \psi}{\psi}$ and the ``quantum velocity'' vector $\frac{2\bm{\nabla}_i \psi}{\psi}$.
The relation
\begin{equation}
\frac{\nabla^2\psi}{\psi} = \nabla^2 \log \psi + (\bm\nabla \log \psi)^2,
\label{eq:DerivativeAsLogDerivative}
\end{equation}
allows us to separate the contributions from the Jastrow and the coordination terms.
The later is labeled below as $\psi_c$.
We also use a label $(\cdot)_{s,t}$ for the $t$-{th} spatial dimension corresponding to particle $s$;
that is, $1\le t\le D$ and $s$ spans from $1$ to $N$.
It is convenient to define vectors $\bm{v}$ and $\bm{u}$,
\begin{align}
v_{s,t} & =  \frac{1}{S_s} \sum_{i\ne s} y_2'(r_{si})\frac{x_{s,t}-x_{i,t}}{r_{si}} , \label{eq:VectorV}\\
u_{s,t} & = \sum_{i\ne s} y_2'(r_{si})  \frac{1}{S_i} \frac{x_{s,t}-x_{i,t}}{r_{si}} . \label{eq:VectorU}
\end{align}
The quantum velocity is obtained by
\begin{equation}
\nabla_{s,t} \log \psi_c = ( \bm{u}+\bm{v} )_{s,t}.
\label{eq:FirstLogDerivative}
\end{equation}
The second derivative, summed on the spatial dimension, can be written as
\begin{multline}
\sum_{t=1}^D \nabla_{s,t}^2 \log \psi_c= 
\sum_{i\ne s} 
\left\{ 
   \left[
      y_2''(r_{si}) + \frac{D-1}{r_{si}}y_2'(r_{si})
   \right] \times
\right.   
    \\
\times
   \left(
      \frac{1}{S_s}+\frac{1}{S_i}
   \right)
  \left. 
 -
   \left[
      \frac{y_2'(r_{si})}{S_i} 
   \right] ^2
\right\}
 - \sum_t v^2_{s,t} .
\label{eq:SecondLogDerivative}
\end{multline}
Notice the cancellation between $v^2$ terms from (\ref{eq:SecondLogDerivative}) and 
(\ref{eq:FirstLogDerivative}) suggested by Eq.~(\ref{eq:DerivativeAsLogDerivative}).

Written in the above form, it is clear that the relevant calculations
involve the order of $N^2$ operations with storage requirement of only the first order in $N$.
The calculation may proceed as follows.
First, one loops through $N(N-1)/2$ pairs of particles and computes the sums $S$. 
Then the loop is repeated, this time summing the contributions to the 
vectors $\bm v$ and $\bm u$ given by Eqs.~(\ref{eq:VectorV}--\ref{eq:VectorU}) and the contribution to the second derivative
given by the first sum on the r.h.s.\  of  Eq.~(\ref{eq:SecondLogDerivative}).
To complete the calculation of the kinetic energy, one needs to perform $N$
additional operations to compute the second sum in the r.h.s.\ of  Eq.~(\ref{eq:SecondLogDerivative})
and to sum the square of the gradient vector according to 
Eq.~(\ref{eq:DerivativeAsLogDerivative})\footnote{Notice that the  $(\bm\nabla \log \psi)^2$ term
in non-linear; contributions from the Jastrow term have to be added before squaring the vector.}.
Thus the computations with the coordinated wavefunction of Eq.~(\ref{eq:CoordinatedWavefunction}) scales only as the second order in the number of particles,
although the usual loop over the particle pairs needs to be repeated twice.

\subsection{The form of the pair and coordination factors \label{sec:Optimization}}

\begin{figure}
\begin{center}
\includegraphics[width=\linewidth]{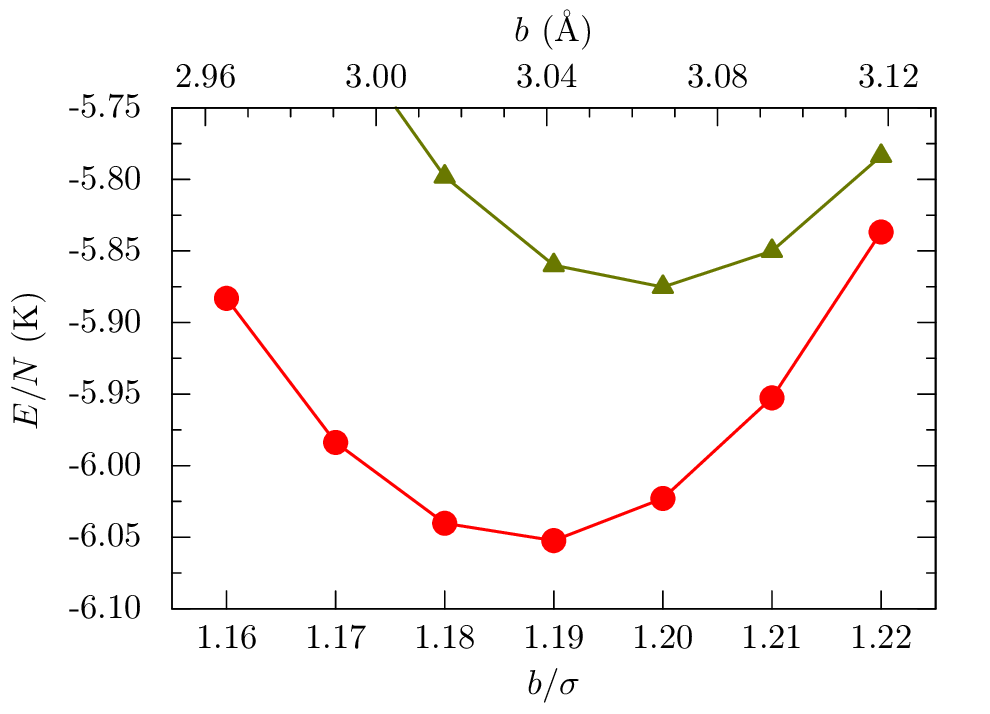}
\end{center}
\caption{ \label{fig:optimization-b-variation} %
Variational energy (per particle) for the coordinated wavefunction  given
by Eqs.~(\ref{eq:CoordinatedWavefunction}), (\ref{eq:ActualWaveFunctionU2})-(\ref{eq:ActualWaveFunctionRr}) (bullets)
and the Jastrow function with the McMillan pair factor given by Eq.~(\ref{eq:ActualWaveFunctionU2}) (triangles).
Both energies are shown as a function of parameter $b$ which enters the two-body correlation factors. 
For each value of $b$, the coordinated function was optimized
with respect to its parameters $m$ and $\delta$.  
For the unit of distance, we use $\sigma=2.556\,$\AA.
The energies were computed for 512-particle systems interacting with Aziz-II\cite{AzizII} pairwise potential.
Statistical errors are smaller than the symbol size.
}
\end{figure}

To test the coordinated wavefunction of Eq.~(\ref{eq:CoordinatedWavefunction}), we have decided to limit the Jastrow term to the simple McMillan form\cite{McMillan} with $u_2=-(b/r)^5$. As this term aims to capture the short-range correlations in the fluid, the mid-range correlations are left to be treated with the coordination term. Having only one variational parameter in the Jastrow product simplifies the  parametrization of the wavefunction. However, the simple form of the McMillan factor misses over 1~K of the correlation energy, most of it due to its imperfection at short distances. One should not hope to recover this energy with any improvement to the mid-range correlations.

For calculations, we used the following form of the pairwise functions,
\begin{align}
u_2(r) & = - \frac{1}{2}\left(\frac{b}{r}\right)^5 - \frac{1}{2}\left(\frac{b}{2 L_c-r}\right)^5  + \left(\frac{b}{L_c}\right)^5 
\label{eq:ActualWaveFunctionU2}
\\
y_2(r) & = 1-\exp\left[ -\left(\frac{\delta}{R(r)}\right)^m \right] 
\label{eq:ActualWaveFunctionG2}
\\
R(r)   & =  \frac{r}{1 - (r/L_c)^4 }
\label{eq:ActualWaveFunctionRr}
,
\end{align}
where $b$, $m$, and $\delta$ are the three variational parameters, and 
$L_c$ is the cutoff distance of the calculation given by half the dimension of the simulation box.


\begin{table}[b] 
\caption{Optimized parameters of the coordinated and non-coordinated Jastrow wavefunctions with the McMillan factor.
Distances are specified in units of $\sigma=2.556$\AA.
Lowest line shows the thermodynamic limit extrapolation of the per-particle energy, with up to 1920 particles used for the calculation.
The interaction was modeled with the Aziz pair potential from Ref.~\onlinecite{AzizII}.
} 
\label{tbl:var-parameters-and-data}
\begin{center}
\begin{tabular}{lclcr}
\hline \hline
&&  Coordinated & \phantom{--} & Non-coord. \\ \hline
$b/\sigma$      &&$1.19$ & & $1.20$  \\
$\delta/\sigma$ &&$1.60$ & & --- \\
$m$             &&$6.55$ & & --- \\ 
\\
$E/N$~(K)       &&  $-6.05(1)$  & & $-5.88(1)$ \\
 \hline
\end{tabular}
\end{center}
\end{table}

The coordination function $y_2$ was chosen to provide a reasonably sharp cutoff beyond a certain distance $\delta$. 
At the same time, we found it quite important to have a ``flat'' $y_2$ at small distances, as 
otherwise the derivative of $y_2$ interferes with the energy terms produced by the derivatives of the 
pair factors $u_2$. 
Because of this effect , using $y_2$ of a Gaussian or exponential form results 
in wide flat energy plateaus in the space of variational parameters. 
Instead, the form given by Eq.~(\ref{eq:ActualWaveFunctionG2}) assures that $y_2$ reaches a constant at small distances.
The relevant small distances are given by the parameter $b$, and the condition can be formulated as
\begin{equation*}
\exp[-(\delta/b)^m] \ll 1 .
\end{equation*}
Satisfying the above condition effectively decouples the optimization of $u_2$ and $y_2$, allowing for a clear interpretation of both terms and for a straight-forward variational optimization. Indeed, we found that variationally optimized parameters $b,\delta$ and $m$ fulfill the above condition to about $10^{-3}$.
 
As is beneficial for a variational calculation, 
both Jastrow and the coordination terms are symmetrized to result in zero gradient 
of the wavefunction at the computational cutoff $L_c$.
The pair term in Eq.~(\ref{eq:ActualWaveFunctionU2}) is symmetrized in the traditional manner, 
while the coordination factor $y_2$ employs
a scaling function $R(r)$ to assure that $y_2$ vanishes smoothly at the cutoff distance $r=L_c$.
The use of the scaling function allows for a robust implementation of the cutoffat $L_c$, yet  introduces
minimal disturbance to $y_2$ at the relevant distances $r\approx\delta$, as in our case $(\delta/L_c)^4 < 10 ^{-2}$.
We found that using scaling function provides a convenient way for symmetrizing the wavefunction.

\section{Results \label{sec:Results}}

\subsection{Variational optimization}

\begin{figure}
\begin{center}
\includegraphics[width=\linewidth]{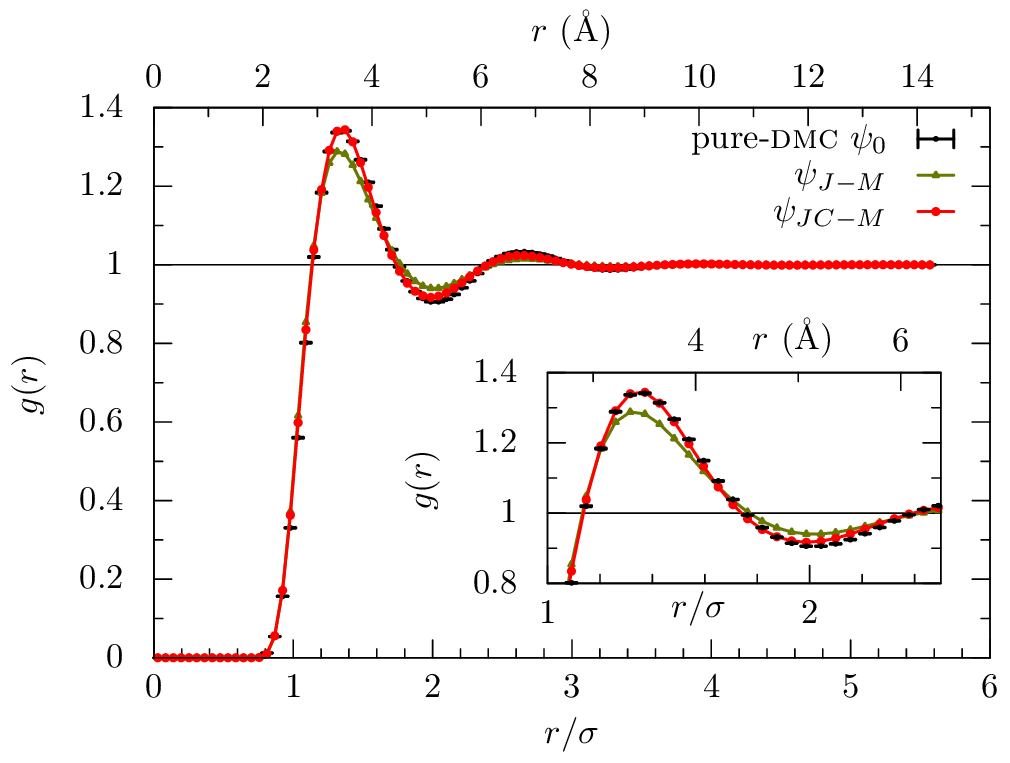}
\end{center}
\caption{
\label{fig:gr}
Pair distribution function $g(r)$ as a function of the interparticle distance, obtained for a 512-atom system.
For the unit of distance, we use $\sigma=2.556\,$\AA.
Unconnected black errors bars: unbiased (pure) estimator obtained with DMC, as described in the text. 
Connected green triangles: Jastrow function with the McMillan pair factor as specified in Eq.~(\ref{eq:ActualWaveFunctionU2}).
Connected red bullets: energy-optimized three-parameter 
coordinated wavefunction $\psi_{JC}$ with the McMillan factor, given by 
Eqs.~(\ref{eq:CoordinatedWavefunction}),(\ref{eq:ActualWaveFunctionU2})-(\ref{eq:ActualWaveFunctionRr}).
Errors bars for both VMC calculations are smaller than their corresponding symbol sizes. 
The inset shows the details of the first correlation peak.
}
\end{figure}

We carried the variational optimization with three-dimensional 512-atom \hefour\ system at the equilibrium density%
\cite{Ouboter1987}
of liquid \hefour, $\rho_0=0.365\sigma^{-3}=21.8~\text{nm}^{-3}$.
Here and below, we use the reduced unit of length equal to $\sigma=2.556\,\text{\AA}$.
All observables where computed on Markov chains generated by the Metropolis 
method \cite{Metropolis-1949} with single-particle updates.
We used a GPU cluster to speed up the calculations using a modification 
of the QL quantum Monte Carlo package \cite{Lutsyshyn2015-FastQuantumMonteCarloOnAGPU}.
The coordinated wavefunction of Eq.~(\ref{eq:CoordinatedWavefunction}) was taken in three-parameters form given by 
Eqs.~(\ref{eq:ActualWaveFunctionU2}--\ref{eq:ActualWaveFunctionRr}).
The system Hamiltonian
\[
\qop{H}=-\frac{\hbar^2}{2m}\sum_i \nabla^2_i + \sum_{i<j} V(r_{ij})
\]
was used with the pair potential by Aziz\cite{AzizII}.

The three variational parameters $b$, $m$, and $\delta$ were optimized on a grid.
Figure~\ref{fig:optimization-b-variation} shows variational energy as a function of parameter $b$, 
given optimal $m$ and $\delta$ for each value of $b$.
The results are compared to the (non-coordinated) Jastrow function with the McMillan factor given by Eq.~(\ref{eq:ActualWaveFunctionU2}).
Optimal value of the parameter $b$ for for the coordinated function was found to be $b=1.19\sigma$, slightly below the optimal value of the non-coordinated 
function, $b=1.20\sigma$. 
As expected, we found little variation in optimal value of $\delta$ with respect to changing the value of parameter $b$. 
In the range shown in Figure~\ref{fig:optimization-b-variation}, optimal $\delta$ varies less than two percent.
%
We also notice relatively weak correlation between parameters $\delta$ and $m$
near the variational minimum.

The optimized values of variational parameters are shown in Table~\ref{tbl:var-parameters-and-data}.
The table also shows the value of optimized energy extrapolated to the thermodynamic limit.
For comparison, Table~\ref{tbl:var-parameters-and-data} also lists the 
energy for the Jastrow function with the McMillan factor.
This energy differs slightly from the one obtained by McMillan\cite{McMillan}, which can be prescribed to the difference in the interaction potential.
As expected, one will notice that the gain in the correlation energy is mild, and amounts to just under 200~mK.
This is in part due to the fact that the missing mid-range correlations are not responsible for a large amount 
of energy, but also because the presence of the coordination term ever so slightly offsets the correlation hole which in turn carries an energy penalty. 

\subsection{Structural properties of the coordinated function}

\begin{table}[b]
\caption{The degree to which the computed pair distribution functions $g(r)$
capture the unbiased estimate $g^*(r)$. The values are computed as $ |g(r_m)-g^*(r_m)|/|1-g^*(r_m)|$, where $r_m$ are the locations of extrema of $g^*(r)$.
The simulation conditions are described in Figure~\ref{fig:gr}.
} 
\label{tbl:captured-g-r}
\begin{center}
\begin{tabular}{lcllllll}
\hline \hline
                &&   $1^\text{st} \max$ &   $1^\text{st} \min$    &  $2^\text{nd} \max$      &  $2^\text{nd} \min$     &   $3^\text{rd} \max$      &  $3^\text{rd} \min$      \\
\hline
Coordinated     && $1.0$ & $0.9$ & $0.7$ & $0.6$ & $0.4$ & $0.4$   \\
Non-coor.       &&  $0.8$  &   $0.6$  &  $0.5$ &  $0.3$  &   $0.2$ &   $0.2$  \\
 \hline
\end{tabular}
\end{center}
\end{table}

\begin{figure*}
\begin{center}
\includegraphics[width=\textwidth]{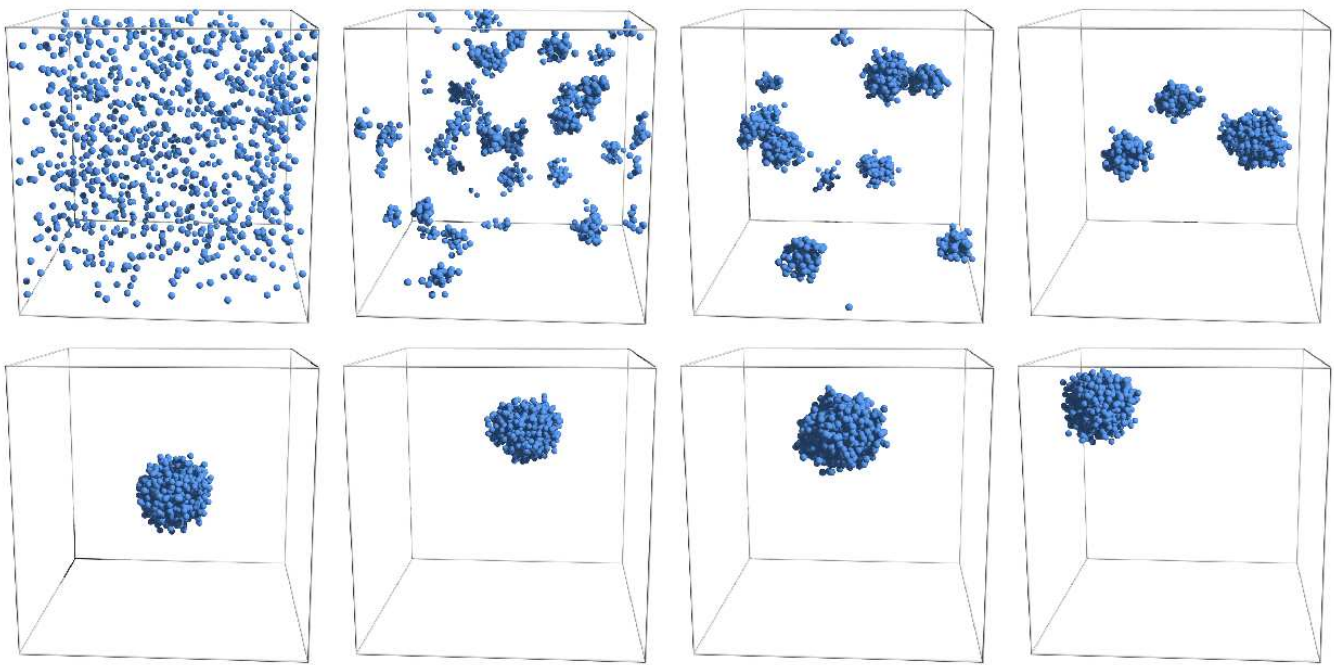}
\end{center}
\caption{
\label{fig:surface-formation}
Progression of Markov chain during Metropolis sampling of a
system with $N=1000$ atoms with the coordinated wavefunction.
The size of the cubic box is equal to $L=100\,\sigma=256\,\text{\AA}$,  
which would correspond to a very dilute homogeneous system.
Top row, left to right, shows the initial state of the system
(with atoms distributed randomly and uniformly), and the system correspondingly after
$10^3$, $10^4$, and $10^5$ macroupdates.
After $10^5$ updates, the Markov chain reaches droplet configuration
which is then sampled throughout the (periodic) simulation cell.
The bottom row, left to right, shows configurations after 
$2\cdot10^5$, $3\cdot10^5$, $10^6$, and $2\cdot10^6$ macroupdates.  
Wavefunction parameters $b=1.19\,\sigma$, $\delta=4.60\,\sigma$, $m=6.55$.
Metropolis sampling was carried via single-particle updates, with
a fixed Gaussian-distribution of displacements 
which resulted in the acceptance ratio of
above 20\% in the homogeneous phase 
to below 35\% in the condensed phase.
Each ``macroupdate'' equals $N$ single-particle Metropolis attempts.
 }
\end{figure*}

As both the potential energy and the wavefunction are built from the 
pairwise functions, the properties of the system are captured 
by the pair distribution function.
The computed pair distribution function $g(r)$ is shown in
Figure~\ref{fig:gr}, along with the results 
for the McMillan function and an unbiased (pure) estimate for the $g(r)$ obtained with the diffusion Monte Carlo (DMC).
The unbiased DMC estimator for $g(r)$ was obtained with the ancestry tracking algorithm of Casulleras and Boronat
\cite{Casulleras3654-UnbiasedEstimatorsInQuantumMonteCarloMethodsApplicationToLiquidHe4}.
Such an unbiased estimator is computed from the projected ground state and can be expected to reflect accurately on the experimental values 
\cite{%
Lester1991-MonteCarloAlgorithmsForExpectationValuesOfCoordinateOperators%
,Casulleras3654-UnbiasedEstimatorsInQuantumMonteCarloMethodsApplicationToLiquidHe4%
,Boronat2005-FreeSurfaceOfSuperfluidHe4AtZeroTemperature%
}.
In properly converged calculations, pure DMC results do not depend on the DMC guiding function.
However, it is worth pointing out that the guiding function for the DMC calculation was  in fact the Jastrow function with the McMillan factor and it did not contain the coordination factor.
In all three cases shown in Figure~\ref{fig:gr}, the calculations were performed
with 512-particle systems at the equilibrium density of \hefour, $\rho=0.365\sigma=21.8\text{ nm}^{-3}$.
The variational parameters were chosen by energy optimization, as specified above, and are given in Table~\ref{tbl:var-parameters-and-data}.

It is notable that the coordinated wavefunction reproduces 
accurately the first correlation peak in the pair distribution function.
The inset in Figure~\ref{fig:gr} shows the detail of the first maximum. 
The first correlation minimum is reproduced slightly less accurately. The following oscillations
in the pair distribution function are also reproduced better by the coordinated wavefunction, albeit with decreasing accuracy.
The position of the maxima and minima in the pair distribution 
was also considerably improved by the coordination term.
The details are given in Table~\ref{tbl:captured-g-r}.
However, the absolute value of these successive oscillations is minute, and they are at distances where the pair potential is vanishing rapidly. 
Thus their influence to the overall energy is nonsignificant.
%


%
%

\section{Inhomogeneous systems \label{sec:SelfBound}}

Jastrow wavefunctions based on a short-range pair factors cannot support the formation of a self-bound state.
That is, a simulation in a sufficiently large box will result in a low-density uniform gas with near-zero potential and kinetic energy. 
Helium liquid, however, is self-bound.
To describe inhomogeneous systems, one generally adds one-body factors 
which bind the liquid phase to a desired shape 
\cite{Schmidt1988,%
   Chester1975-VariationalCalculationsOfLiquidHe4WithFreeSurfaces,%
   Boronat2005-FreeSurfaceOfSuperfluidHe4AtZeroTemperature,%
   Krotscheck1988-StructureAndExcitationsOfLiquidHeliumFilms}. 
This has obvious disadvantages if the surface shape is complex, 
and may pose additional challenges when one needs to maintain the translational symmetry in the system \cite{Lutsyshyn2011-Transmission}. 
Parametrization of the surface adds to the required number of the variational variables.
Self-binding may also be enforced through the use of long-range terms in the two-body
factors, such as introduced in Ref.~\onlinecite{Barranco2012-MgImpurityInHeliumDroplets}, with 
additional term in two-body function $u_2(r)$ proportional to the distance between the particles $r$.
Such a wavefunction serves well as a trial wavefunction for a projector Monte Carlo
calculation, yet variationally, kinetic 
per-particle energy of a system with $u_2\sim-\alpha r$ for large $r$ 
is divergent with the increasing number of particles $N$
as $\sim(\hbar^2/m)\rho_0^{1/3}N^{2/3}\alpha$, where $\rho_0$ is the bulk density.
This presents a number of challenges, as at the very least  $\alpha$ must be $N$-dependent.

The coordinated wavefunction has an unexpected feature
in that by design it supports a self-bound state of the atoms. 
Upon inspection, one will notice that the coordination sum-factors in Eq.~(\ref{eq:CoordinatedWavefunction})
in fact vanish in the limit of low-density, uniformly distributed gas.
Thus the coordination term requires that atoms form clusters, so far as function $y_2(r)$ falls off sufficiently
rapidly with distance%
.
The size
and number of the clusters is determined by the variational parameters and particle density.
For instance, a gas of dimers already  has non-zero coordination term. Increasing the range of $y_2$ 
(which in our case translates to increasing $\delta$ or decreasing $m$) increases the 
size of the clusters. The parameters also provide control over the structure of the surface.

We have carried variational calculation with
the coordinated wavefunction with 1000 atoms 
in a simulation box that resulted in particle density 
$10^{-3}\,\sigma^{-3}\approx0.06\,\text{~nm}^{-3}$.
The Jastrow wavefunction results in a ground state of dilute gas with close to zero energy.
However, the coordinated wavefunction resulted in a bound state
for a wide range of parameters $m$ and $\delta$. Only small $\delta$ 
resulted in the unbound states albeit with positive energy. We also find that the extent of function $y_2$
controls the average cluster size and thus the energy. Variational optimization of $m$ and $\delta$ results
in a state with a single liquid droplet.

To demonstrate the robustness of the inhomogeneous simulation, we carried Monte Carlo 
sampling of the coordinated wavefunction with parameters $m=6.55$ and $\delta=4.50\,\sigma$ (i.e., 
with $y_2$ having larger extent than for the bulk). 
The initial coordinates of the 1000 particles were randomly distributed in the 
simulation box. The sampling sequence is presented in Figure~\ref{fig:surface-formation}.
Soon after the start, the Markov chain arrives at configurations with multiple small clusters. 
As the clusters merge, a single droplet is eventually formed.
The center of mass of the system is not fixed, and the droplet continues to sample 
the entire simulation cell.

The inner structure of the droplets and clusters depend strongly on the two-body function $u_2$.
However, Jastrow function with the McMillan factor 
underestimates the equilibrium density of the bulk \hefour. Without the fixed density
constraint, it is to be expected that the inhomogeneous simulation
should result in lower densities of the condensed phase. 
This was indeed observed.
For example, the droplet shown in Figure~\ref{fig:surface-formation}
has inner density that is less than 70\% of the bulk equilibrium helium density.
Thus the droplet calculation
presented here should be seen as a demonstration of principle. 
The details of their structure, which require a more detailed Jastrow term,  will be the subject of further investigation.

\section{Conclusion}

We have considered a wavefunction ansatz for strongly correlated Bose system that goes beyond the Jastrow-Feenberg expansion. Originating from a symmetrical solid wavefunction
proposed by Cazorla et al.~\cite{Boronat2009NJP}, it is a Bose-liquid wavefunction which explicitly
promotes the creation of the coordination shells around atoms. 
The function is translationally and exchange symmetric. It is fully explicit and is computationally hard as $O(N^2)$, making it well suitable for treatment with quantum Monte Carlo.

To demonstrate the coordination effect, we have studied the wavefunction with the one-parameter McMillan factor for the Jastrow term, and a two-parameter coordination function. The resulting three-parameter wavefunction was straight-forward to optimize variationally. 
The short-range nature of the McMillan factor allowed 
to directly observe the effects of the coordination terms on the mid-range structure of the liquid.
Indeed, the optimized wavefunction results in superior description of mid-range
correlations in the system. 
Comparing with unbiased estimate for the pair distribution function obtained with the diffusion Monte Carlo, we find that the first correlation peak
is reproduced almost exactly.
Moreover, the structure of the pair distribution function is improved consistently throughout 
larger distances as well.

As was first demonstrated in Ref.~\onlinecite{Reatto1974-HowGoodCanJastrowWavefunctionsBeForLiquidHeliumFour},
the first correlation peak can be reproduced rather exactly with the Jastrow function.
However, this required eight variational parameters, and already the description of the first minimum was 
significantly lacking. 
Other approaches to accurately describe the mid-range structure with the Jastrow factors alone have also been reported \cite{ChinUnpublishedPairFunction}.
In our case, the addition of the coordination term allows to separate the short- and middle-range correlations, which can be accounted for correspondingly  by the Jastrow and the coordination terms.

By construction, the coordinated wavefunction supports a self-bound state. 
Consequently, the simulation of inhomogeneous systems does not require the addition of one-body terms.
Moreover, inhomogenuity and surface formation at low densities result directly from the variational optimization of the bulk wavefunction. 
Since the variational ansatz does not require knowledge of the surface geometry, this also provides a powerful tool for cluster states of matter.
However, we find that a satisfactory description of the inhomogeneous phase of helium requires improvements in the Jastrow pair term, which was here limited to the McMillan form for simplicity.
 
The separation of the mid-range correlations into the coordination term which was demonstrated here means that the Jastrow pair term in the coordinated wavefunction only needs to account for the short-range correlations and possibly for the well-understood long-range correlations arising from zero-point phonons. This makes it promising that an accurate short-range pair term can be designed in future with a simple parametrization.


\bibliographystyle{apsrev4-1}


%

\end{document}